\documentclass[twocolumn,showpacs,preprintnumbers,amsmath,amssymb]{revtex4}

\usepackage{graphicx}
\usepackage{dcolumn}
\usepackage{bm}

\begin{document}

\title{The Feynman propagator for quantum gravity:  \\ spin foams, proper time,
  orientation, causality and timeless-ordering}

\author{D. Oriti}
\email{d.oriti@damtp.cam.ac.uk}

\affiliation{Department of Applied Mathematics and Theoretical
  Physics, Centre for Mathematical Sciences, University of Cambridge,
  Wilberforce Road, Cambridge CB3 0WA, UK, EU}

\date{\today}

\begin{abstract}
We discuss the notion of causality in Quantum Gravity in the context
of sum-over-histories approaches, in  the absence therefore of any
background time parameter. In the spin foam formulation of Quantum
Gravity, we identify the appropriate causal structure in the
orientation of the spin foam 2-complex and the data that characterize
it; we construct a generalised version of spin foam models introducing
an extra variable with the interpretation of proper time and show that
different ranges of integration for this proper time give two separate
classes of spin foam models: one corresponds to the spin foam models
currently studied, that are independent of the underlying
orientation/causal structure and are therefore interpreted as a-causal
transition amplitudes; the second corresponds to a general definition
of causal or orientation dependent spin foam models, interpreted as
causal transition amplitudes or as the Quantum Gravity analogue of the
Feynman propagator of field theory, implying a notion of ''timeless
ordering''.   
\end{abstract}

\maketitle

\section{Introduction}
This paper discusses the issue of causality in non-perturbative
quantum gravity, and more precisely in the spin foam formulation of
the theory. This has implications for the broader issue of time in a
background independent quantum theory as well as rather more technical
aspects related to the specific approach we deal with. We will try to
highlight both these conceptual issues and the ideas on which our
results are based and what these results actually are. These results
were presented in \cite{feynman}, and their presentation in section
III, IV amd V is in fact based on \cite{feynman}, and will be analysed
and discussed in more details in \cite{oriented}.

\subsection{Time and causality in Quantum Gravity}
In quantum gravity, because of background independence and
diffeomorphism invariance, neither phsyical observables nor physical
states, i.e. none of the basic elements of any canonical formulation
of the theory, can depend on an external time parameter as is the case
in ordinary quantum mechanics, but time itself becomes a quantum
variable insofar as the metric itself is a quantum variable; this is
true also in a covariant or sum-over-histories formulation, because,
even if the boundary may be assigned timelike data, still the
spacetime geometry cannot be used to define any unique notion of time
in the interior of the manifold as it represents the quantum field we
are summing over to define our theory. This led to the conclusion that
time can not be a basic ingredient in a proper formulation of Quantum
Gravity, or that the very concept of time is not fundamantal at all
and should instead emerge only in a semiclassical approximation
\cite{IshamTime}. Regarding the issue of causality, there seem to be
two basic attitude one can take: one can hold that we cannot speak of
causality in absence of time and therefore just as time itself also a
notion causality should emerge and be applicable only in a
semi-classical limit; this is actually true without argument, actually
a truism, if we stick to the conventional notion of causality in terms
of a given spacetime geometry, and of lightcone structures in a
continuous manifold; however one can argue that the notion of
causality is actually more primitive than that of time, more
fundamental, being already present in the notion of ordering between
events, of a fundamental directionality in spacetime, and that as such
can be present, as a {\it seed} of what will then be, in a
semiclassical approximation, the usual notion of causality, even if
discrete or combinatorial or algebraic structures are used to encode
gravitational degrees of freedom and thus spacetime geometry in a
formulation of Quantum Gravity. We agree with this latter point of
view.

\subsection{Causality as ordering in Quantum Gravity}         
Indeed, certain approaches to Quantum Gravity even take this notion of
causality as order as the basic ingredient for their constructions,
the keystone on which to found the theory; it is the case of the
causal set approach \cite{Sorkin} where spacetime at the most
fundamental level is taken to {\it be} a set of fundamental events
endowed with an ordering representing their causal relations and the
task is that of constructing a suitable quantum amplitude for each
configuration so specified, this amplitude being the main ingredient
ina sum-over-histories formulation of quantum gravity as a sum over
causal sets, i.e. a sum over causally ordered or oriented, discrete
sets of events. 

This idea of causality as ordering as a basic ingredient in a
sum-over-histories formulation of the theory has a formal
implementation also in the analytic path integral approach to quantum
gravity as developed by Teitelboim \cite{teitelboim}, so not assuming
any discrete substratum for spacetime, and his construction can be
seen as a formal realization of what seems to be a general principle
in any sum-over-histories quantum gravity theory: implement the
causality principle by requiring that each history summed over,
i.e. each spacetime geometry, is {\it oriented}, that this orientation
matches that of the boundary states/data and is registered in the
quantum amplitude assigned to each history, such that, when a
canonical formulation is available, the two boundary states can be
distinguished as either 'in-state' or 'out-state'; this orientation
dependence replaces the notion of 'time-ordering' of quantum field
theory, that is of course unavailable in this background independent
context. The fact that this is possible or not, thus the exact way the
Quantum Gravity transition amplitudes are defined allows a distinction
between different kinds of transition amplitudes, causal and a-causal
ones. More precisely \cite{teitelboim}, the distinction between the
different transition amplitudes in registering initial and final
3-geometry $(h^1,h^2)$ in the boundary data of the path integral is
obtained by choosing different range of integration in the proper
time formulation of the theory; using a canonical decomposition of
the metric variables, the role of proper time is played by the
lapse function; an unrestricted range of integration leads
to the (quantum gravity analogue of the) Hadamard function, while
a restriction to positive lapses leads to the (analogue of the)
Feynman propagator. The expressions for the two quantities have
just a formal meaning, but read: 
\begin{eqnarray}
G_H(h^2,h^1)=\int_{-\infty}^{+\infty}\mathcal{D}N\left[
e^{i \int_{\mathcal{M}} d^3x dt \left(\pi^{ij}\dot{h}_{ij} - N
\mathcal{H} - N^i \mathcal{H}_i\right)} \mathcal{D}\pi
\mathcal{D}h\right] \nonumber \\
G_F(h^2,h^1)=\int_{0}^{+\infty}\mathcal{D}N\left[ e^{i
\int_{\mathcal{M}} d^3x dt \left(\pi^{ij}\dot{h}_{ij} - N
\mathcal{H} - N^i \mathcal{H}_i\right)} \mathcal{D}\pi
\mathcal{D}h\right] \nonumber. 
\end{eqnarray} 
The idea behind this construction is best understood recalling basic
facts from the theory of a quantum relativistic particle (that cna be
thought indeed as quantum gravity in $0+1$ dimensions. There
\cite{halliwell} the main difference
between the different transition amplitudes (and 2-point
functions) one can construct, in particular the Feynman propagator
and the Hadamard function, is exactly the way they encode (or fail
to encode) causality restrictions, in the fact that the amplitude
registers or not that one of the two is in the causal future of the
other; this order can be
imposed in each of the histories summed over in a clear way using
a proper time formulation: starting from the same proper time
dependent expression $g(x_1,x_2,T)$, which can be given a
sum-over-histories form, the Hadamard function is obtained
eliminating this dependence by integrating over the proper time
variable with an infinite range over both positive and negative
values; the resulting amplitude is then a-causal, and real, and
does not register any ordering between its arguments; the Feynman
propagator is instead obtained restricting this integration to
positive or negative proper times only, and this ordering is
precisely what makes it {\it causal}; denoting
$G_W^{\pm}(x_1,x_2)$ the positive and negative Wightman function,
the result is: 
\begin{eqnarray}
&&G_H(x_1,x_2)=G_W^+(x_1,x_2)+G_W^-(x_1,x_2)= \nonumber \\
&&= \int_{-\infty}^{+\infty}dT \,\,g(x_1,x_2, T) \nonumber \\ \nonumber
\\ && G_F(x_1,x_2)=\theta(x_2^0 - x_1^0)\,G_W^+(x_1,x_2)+ \nonumber \\
&&+\theta(x_1^0 -
x_2^0)\,G_W^-(x_1,x_2)=\int_{0}^{+\infty}dT\,\,g(x_1,x_2,T)
\nonumber \\ \nonumber \\ &&g(x_1,x_2,T)=\int d^4 p \,\,e^{i [(x_2-x_1) \cdot p -
T (p^2 + m^2)]} \nonumber. 
\end{eqnarray}
The problem we address in
this work is how to construct the analogue quantities in a spin
foam context, i.e. in a purely combinatorial, algebraic and group
theoretic way, in absence of any smooth manifold structure and any
metric field. We will see that this can be achieved in full
generality and in a very natural way.

\section{Spin Foam models of Quantum Gravity}
Spin foam models \cite{review, perez} are currently being studied
as a new more rigorous implementation of the path integral
approach to quantum gravity. As such they are constructed by a
definition of histories of the gravitational field, interpreted as
4-dimensional geometries for a given spacetime manifold, and an
assignment of quantum amplitudes to these geometries, i.e.
suitable complex functions of the geometric data characterizing
each history. Different models have been proposed and derived from
many different points of view, including lattice gauge theory type
derivations \cite{danruth} and group field theory formulations
\cite{alecarlo}, for both the Riemannian and Lorentzian signatures,
and in different dimensions, which counts as one of the
attractive features of this approach. The peculiarity of the spin
foam framework, as compared with the traditional path integral for
gravity, is that the spacetime manifolds on which the
gravitational data are given are combinatorial and discrete ones, 
and specifically are given by combinatorial 2-complexes,
i.e. collections of vertices, edges and faces together with their
relations (\lq\lq what is in the boundary of what\rq\rq ), and the
histories of the gravitational field are characterized by data
taken uniquely from the representation theory of the local gauge
group of gravity, i.e. the Lorentz group, and no familiar notions
of metric on differentiable manifolds are used. While this maybe
very attractive at the aesthetic/philosophical level, and turns
out to be very useful at the technical level (it is because of
this that precise definitions of both the measure and the
amplitudes for 4-geometries can be given), it makes the resulting
models more difficult to interpret and to work with, as we lack
all our conventional GR tools and quantities, as well as our geometric
intuition. The main object that such a sum-over-histories formulation
defines is the partition
function of the theory, and from this, allowing the underlying
spacetime manifold (2-complex) to have boundaries, quantum
amplitudes functions of the boundary data, that when a canonical
interpretation is available should be interpreted as transition
amplitudes between quantum gravity states. The partition function
in these models, for a given 2-complex $\sigma$, has the general
structure: 

\begin{equation}
Z(\sigma)=\sum_{\{\rho_f\}}\prod_f\mathcal{A}_f(\rho_f)\prod_e\mathcal{A}_e(\rho_{f\mid
  e})\prod_v\mathcal{A}_v(\rho_{f\mid v}) 
\end{equation} 
where the amplitudes
for faces $f$, edges $e$ and vertices $v$ of the 2-complex are all
functions of the representations $\rho_f$ of the Lorentz group
associated to the faces of the complex.

The analysis we are going to describe in the following concerning the
role of orientation data in the spin foam context and the
implementation of causality is
restricted to models based on ordinary Lie groups and homogeneous
spaces, like the Ponzano-Regge models \cite{freidel} in 3 dimensions
and the Barrett-Crane type models in 4 and higher, and not using
explicitely any quantum group structure.
Physically these can be interpreted as models of quantum gravity
without cosmological constant. We make use of expressions for the
amplitudes of the models in terms of both group variables (or
variables with values in an homogeneous space) and representation
variables, i.e. we use a 'first order' formulation of spin foam
models \cite{lo}. We do not consider less understood spin foam
models, e.g. the Reisenberger model, although it is quite likely,
in our opinion, that a similar analysis can be performed also in
that case. For the rest our analysis is completely general and
holds in any spacetime dimension and any signature
\cite{feynman, oriented}. For the sake of brevity, however, we will
show how the construction works more
explicitely only for the Barrett-Crane type models in $n$-dimensions
based on the homogeneous space $H^{n-1}\simeq SO(n-1,1)/SO(n-1)$ or
$S^{n-1}\simeq SO(n)/SO(n-1)$.
In these models, the
amplitudes in the 1st order formalism factorize for the different
vertices of the 2-complex \cite{review} and take the form:

\begin{eqnarray}
Z=\left(\prod_f \int d\rho_f\right)\left(\prod_v \prod_{e\in v}
\int_{H_e}dx_e \right)\prod_f \mathcal{A}_f(\rho_f) \nonumber \\
\prod_e\mathcal{A}_e(\rho_{f\in e})\prod_v\mathcal{A}_v(x_{e\in
  v},\rho_{f\in v})
\end{eqnarray}
where the 2-complex is taken to be topologically dual to a
$n$-dimensional simplicial complex (with a $1-1$ correspondence
between $k$-cells of the 2-complex and $n-k$-simplices of the
simplicial complex) and therefore the precise combinatorics varies according to the
dimension, but in any case: $\rho_f$ are the unitary irreps of the
local gauge group of gravity ($SO(n-1,1)$ in the n-dimensional
Lorentzian case and $SO(n)$ in the n-dimensional Riemannian case), and
these unitary representations are labelled by either a
half-integer in the Riemannian case and in the case of discrete simple
representations $(n,0)$ of $SO(n-1,1)$, or by a real parameter in the
case of continuous simple representations $(0,\rho)$ of $SO(n-1,1)$,
$H_e$ is the homogeneous space to which the vectors $x_e$ belong and
is $S^n$ in the Riemannian models and the $n-1$-dimensional
hyperboloid $H^{n-1}\simeq SO(n-1,1)/SO(n-1)$ in the
Lorentzian models we consider here. The algebraic data $\{\rho_f\}$
and $\{ x_e\}$ have a geometric interpretation in that the $\rho_f$
are to be thought of as volumes of the $(n-2)$-simplices dual to the
faces of the 2-complex and the $x_e$ are to be thought of as unit normal
vectors to the $(n-1)$-simplices dual to the edges of the 2-complex. 
While the expression for the edge amplitudes, interpreted as part of
the gravitational measure together with the face amplitudes, varies in the various
models, the vertex amplitude is the same in all these type of models
and taken to be:
\begin{eqnarray}
\mathcal{A}_v(x_{e\in v},\rho_{f\in v})=\prod_{f\in
    v}\mathcal{A}_{f\in v}(x_{e1\in f},x_{e2\in f},\rho_f)= \nonumber
\\ =\prod_{f\in v}\mathcal{A}_{f\in v}(\theta_f,\rho_f)=\prod_{f\in
  v} D^{\rho_f}_{00}(\theta_f)
  \end{eqnarray}
i.e. they are given in terms of zonal spherical functions $D_{00}^{\rho_f}(\theta_f)$ of the Lorentz group in the representation $\rho_f$ \cite{V-K}, where we have indicated that the amplitudes depend on the variables on
the homogeneous spaces only through the invariant distances
$\theta_f=x_{e1\in f}\cdot x_{e2\in f}$ between the vector associated
to the two edges in the boundary of each face that are inside each
vertex. 

The structure is therefore that of
a discrete path integral for gravity with a combinatorial structure
playing the role of the base spacetime manifold, algebraic data living
on it and playng the role of the gravitational degrees of freedom, and
a (precisely defined) quantum amplitude and measure assigned to each
configuration. In the group field theory approach to spin foam models
(see \cite{review,perez}), a sum over 2-complexes is also beautifully
implemented. The issue now is to show that, on the one hand, it is
possible even in this purely combinatorial/algebraic context to
identify a notion of causal structure and appropriate data
representing it, and then, on the other hand, that the spin foam
formalism is flexible enough to provide us with a definition of both
causal and a-causal transition amplitudes for quantum gravity, as the
usual path integral formalism does for quantum particle dynamics or
quantum field theory.

\section{Causality as orientation in spin foam models}
As we said, in spin foam models spacetime is replaced
by a combinatorial 2-complex. Extra
data would then assign geometric information to this spacetime
structure. What can be the analogue of causal relations in such a
context? Where is causality to be looked for? In the end the
problem is solved more easily than could a priori be expected. Consider
just the first layer of the spin foam 2-complex, i.e. only
vertices and links connecting them. This is basically just a
graph. If we add to it orientation data, i.e. arrows on the links,
we obtain an oriented (or directed) graph, a set of oriented links connecting a
set of vertices. Now the vertices can be interpreted as a set of
fundamental spacetime events and the oriented links are then the
causal relations between them (this is also consistent with an
'operational interpretation' of the dual simplicial structure, with
$n$-simplices representing inperfect, thus realistic, definition of
spacetime points/events). We can assign an {\it orientation
variable} $\alpha_{e\mid v}$ to each link, with respect to each
vertex $v$ it connects, that takes the values $\pm 1$ according to
which orientation is chosen. This auotmatically defines 'oriented
normals' $n_e=\alpha_{e\mid v} x_{e\mid v}$ from the un-oriented ones
$x_e$. At the same time we can assign
another orientation variable to each vertex, call it $\mu$ again
taking the values $\pm 1$. The spacetime interpretation of these
variables is that of indicating whether the vertex is a future
pointing or past pointing contribution to the overall spacetime
diagram, and clearly a positively oriented link with respect to a
future pointing vertex is equivalent to a negatively oriented one
with respect to a past oriented vertex; this means that what gives
the spacetime orientation of each link in each vertex is actually
the combination $\alpha_{e\mid v}\mu_v$. A consistency condition
for the assignment of orientation data to the graph is that when a
link $e$ connects two vertices it has the opposite orientation in
the two, which has a clear spacetime interpretation: 
$
\alpha_{e\mid v_1}\mu_{v_1} = - \alpha_{e\mid v_2}\mu_{v_2}
\label{orientation1}$. 
Taking now into account the full
combinatorial structure of the 2-complex, we also assign an extra
orientation variable to each face $\epsilon_f=\pm 1$. It is
crucial to notice that the structure we have been describing is
basically that of the Hesse diagram representing a causal set, but fails
to be that of a proper poset or causal set \cite{Sorkin} because
the set of vertices-events endowed with the ordering relation
represented by the arrow fails to satisfy in general any
antisymmetry condition, i.e. it is generally not the case that following the links according to their
orientation we never end up at the starting point; in other words,
our causal relations allow for closed timelike loops. Also, notice
that this causal interpretation of our oriented (or directed)
graph makes sense only in a Lorentzian context, when the signature
allows for a spacetime translation in terms of lightcones;
however, the structure we have been describing remains the same
even if thought of in a Riemannian signature, and in fact the
issue we will be confronting in the following is the general one
of constructing spin foam models that reflect and take into
account appropriately the {\it orientation} of the underlying
2-complex, i.e. of {\it orientation-dependent transition
amplitudes} for quantum gravity. Only in a Lorentzian context
these will have the interpretation of causal amplitudes or of
quantum gravity analogues of the Feynman propagator. The
2-complexes used in spin foam models are not generic: they are
topologically dual to simplicial n-dimensional manifolds: to each
vertex corresponds a n-simplex, to each link a (n-1)-dimensional
simplex, to each face an (n-2)-dimensional simplex; note that this gives another restriction with respect to a generic causal set. The
orientation data we assigned have then a clear geometric
interpretation in this simplicial picture: the $\mu_v$ variable
for a vertex takes the values $\pm 1$ according to whether the
n-simplex dual to it is isomorphic to a n-simplex in Minkowski
(Euclidean) space or the isomorphism holds for the opposite
orientation; the variable $\alpha_{e\mid v} =\pm 1$ indicate
whether the normal to the (n-1)-simplex dual to the link $e$ is
ingoing or outgoing with respect to the n-simplex dual to $v$, and
the variables $\epsilon_f$ also characterize the orientation of
the $n-2$-simplex dual to the face $f$. Knowing this dual
geometric interpretation of the elements of the 2-complex, it is
easy to derive a consistency condition on the values that these
orientation variable must take to correspond to a well-posed
orientation of the 2-complex (simplicial manifold); the relation, that
basically
follows from Stokes's theorem \cite{lo} is: 
\begin{equation}
\forall v \;\;\epsilon_{f\mid v} = \alpha_{e_1 \mid
v}\alpha_{e_2\mid v}\mu_v \label{oricond},
\end{equation} 
where $e_1$ and $e_2$ label the two
links that belong to the boundary of the face $f$ and touch the
vertex $v$. 

This orientation structure is what we can identify at
the quantum level as the seed for the emergence of causality in
the classical limit, in a Lorentzian context, as we said; now the
question is: do current spin foam model take this into account in
their amplitudes? The answer \cite{feynman,oriented} is that all
current spin foam models do not depend,
in their amplitudes, on the orientation of the underlying 2-complex,
i.e. they do not depend on the orientation data we identified above.

The way this is achieved is quite simple in all models: in the
expression for the amplitudes for spin foams the terms that can be
understood as contributions from opposite orientations are summed
simmetrically thus erasing the dependence on the orientation
itself. 
In the Barrett-Crane type of models, for example, the orientation independence is achieved at the level of each face in each vertex:

\begin{eqnarray}
\lefteqn{\mathcal{A}_v(x_{e\in v},\rho_{f\in v})=\prod_{f\in v}\mathcal{A}_{f\in v}(\theta_f,\rho_f)=} \nonumber \\ &=&\prod_{f\in
  v}\left( \mathcal{W}_f^{\epsilon_f=+1}(\theta_f,\rho_f) +
\mathcal{W}_f^{\epsilon_f=-1}(\theta_f,\rho_f)\right)= \nonumber \\
&=&\prod_{f\in v}\left(
\mathcal{W}_f^{\mu_v=+1}(\alpha_{e1}\alpha_{e2}\theta_f,\rho_f) +
\mathcal{W}_f^{\mu_v=-1}(\alpha_{e1}\alpha_{e2}\theta_f,\rho_f)
\right)\;\;\;\; \label{indep2}
\end{eqnarray}
where we have indicated that the amplitudes depend on the variables on
the homogeneous spaces only through the invariant distances
$\theta_f=x_{e1\in f}\cdot x_{e2\in f}$ between the vector associated
to the two edges in the boundary of each face that are inside each
vertex, and in the last step we have traded the orientation data on
the faces for those on the vertices by using relation
\ref{oricond}. 
The mathematics behind this structure and the nature of the
orientation-{\emph dependent} functions $\mathcal{W}$ is very general, i.e. independent of signature and dimension, and it is explained in \cite{feynman,oriented}.

This orientation independence leads to interpreting the current spin
foam models as a-causal transition amplitudes, as we said, i.e. as the
quantum gravity analogue of the Hadamard function of field theory and
particle dynamics; when a canonical formulation is available, this
sort of transition amplitudes can be equivalently thought of as
defining the physical inner product between quantum gravity states,
invariant under spacetime diffeomorphisms, or as a covariant
definition of the matrix elements of the projector operator onto
physical states \cite{carloprojector}. 

Given the universal structure outlined above, we are lead to look for
a universal way of modifying current spin foam models and to a new,
again universal, definition of orientation-dependent or causal spin
foam models \cite{oriented}; these would correspond to causal
transition amplitudes and to a quantum gravity analogue of the Feynman
propagator of field theory. The \lq\lq brute force\rq\rq way is obvious and was
performed in the 4-dimensional Lorentzian case in \cite{lo}: 1) Impose
consistency conditions on orientation parameters $\epsilon_f$,
$\alpha_e$ and $\mu_v$; 2) choose the orientation of simplices, in
particular the value of $\mu_v$ (future or past-pointing), 3) restrict
the spin foam amplitudes to include only the chosen orientation,
i.e. switch from D functions to $W$ functions dropping the sum over
$\mu_v$ or $\epsilon_f$. Hovewer, we will shortly see that a much more
natural and elegant construction exists, that makes use of a
generalised formulation of spin foam models in terms of a new proper
time variable, from which all the different types of transition
amplitudes can be defined, just as in the particle/QFT case. Also, the
expression
\ref{indep2} is reminescent of the decomposition of Hadamard
functions for relativistic particles or fields into Wightman
functions, in turn the basic elements in the definition of the Feynman
propagator. The construction we are going to describe now takes this analogy
seriously and shows that it is indeed exact, and uses it as the
staring point for a spin foam definition of the quantum gravity
Feynman propagator that implements a notion of 'timeless ordering'.

\section{Particles on Lie groups/homogeneous spaces}
The amplitudes assigned to spin foam faces, edges and vertices in the
Barrett-Crane-type models is given by the evaluation of simple spin
networks and was described in \cite{FK} in analogy to the evaluation
of Feynman diagrams: 1) assign a variable valued in the relevant
homogeneous space to each vertex of the given spin network, 2) assign a
zonal spherical function $D^{\rho_f}_{00}(\theta_f)$  to each line and
3) sum over all the possible values of the variables on the vertices to
get the final amplitude. In this prescription the zonal spherical
function is treated as a kind of propagator, and indeed on the one
hand it turns out
that this feynmanology has its roots in a (group) field theory of
these models \cite{review}, and, on the other hand, it can be shown
\cite{feynman, oriented} that the zonal spherical function used is
indeed the Hadamard function for a scalar particle on the
homogeneous space on which the models are based. This fact prompts a
completely general definition of generalised, first, and then causal
spin foam models.

Consider a scalar field $\phi(g)$ with mass $m$ living on the Lie
group $G$ or on the homogeneous space $H$, with each point on it
labelled by $g$; consider its free evolution parametrised by a proper
time coordinate $s$; the equation of motion in proper time is: $(i
\partial_s + \Delta) \phi(g,s) = 0$ with $\Delta$ being the
Laplace-Beltrami operator $G$ ($H$). The dynamics is completely
captured by the evolution kernel $K(g,g',s)=K(g g^{-1},s)$
\cite{camporesi}, in the sense that given the initial condition
$\psi(g_0,0)$, we have: $\psi(g,s)=\int dg_0
K(g,g_0,s)\psi(g_0,0)$. The dependence on the proper time variable
should of course be eliminated and the various physical propagators
are obtained from the evolution kernel according to how this is
accomplished; the Hadamard function is obtained via the expression:
$H(g,g',m^2)= -i \int_{-\infty}^{+\infty}ds K(g,g',s)e^{-i m^2 s}$,
while restricting the range of integration to positive proper times
only gives the feynman propagator $G_F(g,g',m^2)= -i \int_0^{+\infty}
ds K(g,g',s) e^{-i m^2 s}$, where the usual Feynman prescription for
the contour of integration ($m^2 \rightarrow m^2 -i\epsilon$) is
assumed for reasons of convergence (one may look at the two
expressions as resulting from either a Fourier or a Laplace transform
of the same function of proper time \cite{oriented}).    

It turns out \cite{oriented} that the functions entering the
expressions for the quantum amplitudes of all current spin foam models
correspond to the Hadamard 2-point functions for a scalar field on the
relevant Lie group/homogeneous space with $m^2=-C(\rho_f)$ where
$C(\rho_f)$ is the Casimir eigenvalue of the simple irreducible
representation labelling the face of the 2-complex (therefore the link
of the spin network whose evaluation gives the amplitude for the
vertex):

\begin{eqnarray}   
&H(\theta_f, m^2)=-i\int_{-\infty}^{+\infty}ds
  K_{\mathcal{M}}(\theta_f,\mu_v s)e^{+iC(\rho_f)\mu_v s}= \nonumber
  \\ &=\frac{i}{2\pi}\sqrt{\Delta_{\rho_{f\in v}}} D^{\rho_{f\in
      v}}_{00}(\theta_{f})&,
\end{eqnarray} 
where the relevant orientation data $\mu_v$ enter in the definition of
the proper time variable being integrated over, $\Delta_{\rho}$ is the
dimension of the representation $\rho$ in the Riemannian case, or in
the Lorentzian case (where the unitary representations are infinite
dimensional) the contribution of the representation to the Plancherel
measure, and the Casimir eigenvalues are $C(\rho)=2j(2j+n-2)$ with $j$
half-integer in the Riemannian case, and $C(\rho)= -\rho^2 -(n-2/2)^2$
with $\rho$ positive real in the Lorentzian case based on the timelike
hyperboloid.
It is clear that the result is independent of the value of the
various orientation data.

On the other hand simply imposing the above restriction in the proper
time integration (that amount indeed to a causality restriction for
the particle evolution \cite{teitelboim}), one obtains the expression
for the Feynman propagator that one needs to define orientation
dependent or causal spin foam models, that turns out \cite{oriented}
to have the expected expression in terms of $W$ functions
confirming the interpretation of these as
Wightman functions:
\begin{eqnarray}
\lefteqn{G(\theta_f, m^2,\mu_v)=-i\int_{0}^{+\infty}ds
  K_{\mathcal{M}}(\vartheta_f,T)e^{+iC(\rho_f)T}=} \nonumber \\
 &=\frac{i}{2\pi}\sqrt{\Delta_{\rho_{f}}}\left[
  \theta(\mu_v)\mathcal{W}^{+}(\vartheta_{f},\rho_f)+\theta(-\mu_v)\mathcal{W}^{-}(\vartheta_{f},\rho_f)
  \right]\,\,\,\,\,\,,
\end{eqnarray}
where we have denoted with $\vartheta$ the \lq\lq oriented angle\rq\rq
$\vartheta_f=\alpha_{e1\in f}\alpha_{e2\in f}\theta_f$, and with the
\lq\lq oriented proper time\rq\rq being $T=\mu_v s$. Similarly for the
other spin foam models \cite{oriented}.

This is a non-trivial function of the orientation data, with the usual
\lq\lq time ordering\rq\rq being replaced by a \lq\lq timeless
ordering\rq\rq !

\section{A proper time formalism for spin foam models and the spin
  foam definition of the Feynman propagator for quantum gravity}

Taking seriously this particle analogy and the associated spin network
feynmanology, having also in mind their group field theory formulation
\cite{review}, we can generalise the current formulation of spin foam
models to include a proper time variable, and obtaining a general
expression from which different transition amplitudes and consequently
both orientation independent and causal models can be derived, simply
changing the integration contour in proper time. The formulae have of
course to include an explicit dependence on the orientation data. The
expression for the n-dimensional spin foam models ($n>3$) based on the
homogeneous space $\mathcal{M}$ looks as follows:

\begin{eqnarray} 
\lefteqn{Z=\left( \prod_v \prod_{f\mid v}\int_C ds_{f}\right) \left(
  \prod_f \sum_{\rho_f}\right)\left( \prod_v \prod_{e \mid v}
  \int dx_e \right) {\large \mathcal{A}(\rho, x,
    T)}} \nonumber \\ \nonumber \\ &{\large \mathcal{A}(\rho,
  x, T)}= \prod_f \mathcal{A}_f(\rho_f) \prod_e
\mathcal{A}_e(\rho_{f\mid e})\;\;\;\;\;\;\;\;\;\;\;\;\;\;\;\;\;\;
\nonumber \\ &\prod_v\prod_{f\mid v} \left( -2\pi i
(\sqrt{\Delta_{\rho_f}})^{-1} \,\,K(\vartheta_{f\mid v},T_{f\mid v}) e^{i
  C(\rho_f) T}\right). 
\end{eqnarray}
A similar formula holds for the 3-dimensional models.
This generalised expression encompasses both the usual un-oriented
models and the new causal ones; indeed the first are obtained by
choosing the extended range of integration $C=(-\infty, +\infty)$ for
the $s$ variable (and this erases the dependence on $\mu_v$ and
$\alpha_e$), while the quantum gravity Feynman propagator is obtained
with $C=(0,+\infty)$. In this last case, a regularization prescription
for convergence is implicit for each variable (so that the expression
has to be understood in the complex domain): $\rho_f \rightarrow
\rho_f +i \epsilon$, $\vartheta_f \rightarrow \vartheta_f + i
\delta$. The explicit form of the evolution kernel $K$ differs of
course in the different models and affects the exact form of the
amplitudes \cite{oriented}, that however all share the general
structure here presented.

Notice that we have not modified the face and edge amplitudes with
respect to the usual models. We could have done it: these amplitudes
admit a Feynman graph-like evaluation as well and the technology
related to quantum particles on homogeneous spaces could have been
used to generalise and then modify them as we have done for the vertex
amplitudes. The reason why we have not done so is twofold: on the one
hand the usual interpretation of these contributions to the spin foam
models is that of a conribution to the overall measure, therefore the
implementation of causality is needed only at the vertex level, that
is instead supposed to encode the dynamics of the theory; on the other
hand, the form of the edge amplitudes in some version of the 4-d
Barrett-Crane model is understood as arising directly from the form of
boundary spin network states \cite{boundary}, and we prefer, at this
stage of development of the theory, to keep that structure without
modification. We believe, however, that alternative formulations of
the models and possible definitions of modified spin network states,
maybe in order to induce an orientatation-dependence in their
structure, deserve further analysis.   
 
Let us discuss the various properties of the new kind of transition
amplitudes for quantum gravity we are defining here, the causal or
orientation dependent spin foam models. 

The first property we would like to stress is the fact that all these
models can be recast in the form of quantum causal histories models
\cite{fotini} as it was done for the 4-dimensional Lorentzian case in
\cite{lo}; therefore they define highly non-trivial amplitudes for
causal sets, if the combinatorial structure of the underlying
2-complex is such that it does not contain closed timelike loops,
although they are ceratainly a restriction of the possible causal sets
one can consider since the vertices are here restricted to be
$(n+1)$-valent in n dimensions. 

Second, the quantum amplitudes defined by these models can be related
very easily to classical simplicial gravity, of which they clearly
represent a covariant quantization: while the un-oriented models can
be related to the classical Regge action only in a asymptotic limit,
when their vertex amplitudes result in being proportional to the
cosine of it, here the connection with the Regge action is manifest;
for example, in the 4-dimensional case the relevant evolution kernel
has the form \cite{camporesi}: 

\begin{equation}
K(\vartheta,T) = \frac{1}{(4\pi i T)^{3/2}}\left(
\frac{\vartheta}{\sinh\vartheta}\right) e^{i\frac{\vartheta^2}{4T}
  -iT} 
\end{equation} 
the causal vertex amplitude, after the (restricted) proper time
integration is perfomed, takes the form:
\begin{eqnarray}
\lefteqn{\mathcal{A}^C_v = \prod_{f \in v} \left( -2\pi i \rho_f
  \int_0^{+\infty} ds_f K(\vartheta_f,T) e^{i C(\rho_f)T}\right)
  =\;\;\;\;} \nonumber \\ &=& - \left( \prod_f
\frac{1}{\rho_f\sinh\vartheta_f}\right) \,e^{i\sum_{f\in v} \mu_v
  \rho_f \vartheta_f}\;\;\;\;\;\;\;\;\;\;\;  
\end{eqnarray}  
and therefore the product over vertex amplitudes in the spin foam model gives
\begin{eqnarray}
\prod_v\mathcal{A}^C_v=\left(\prod_v\prod_{f\in
  v}\frac{1}{\rho_f\sinh\vartheta_f}\right)\,e^{i\sum_{f\in v} \rho_f
  \sum_v \mu_v  \vartheta_{f\in v}}= \nonumber \\ =
\left(\prod_v\prod_{f\in
  v}\frac{1}{\rho_f\sinh\vartheta_f}\right)\,e^{i {\Large
    S_{R}}(\rho_f,\vartheta_f)}\;\;\;\;\;\;\;\;,
\end{eqnarray}
i.e. the exponential of the Regge action (in first order formalism
\cite{lo}) for simplicial gravity, apart from an additional
contribution to the overall measure, as one would expect from a
sum-over-histories formulation of quantum gravity based on a
simplicial discretization. The same can be shown to be true
explicitely in the 4d Riemannian case and in the 3-dimensional case,
while the proof for the higher dimensional models is made more
complicated by the lack of a simple enough expression for the
evolution kernel \cite{oriented}.

\section{Conclusions}
Let us summarise what we have presented. We have discussed in which
sense causality may be thought of pre-existing (!) time at the most
fundamental level, and what notion of causality, interpreted as a
'seed' from which the usual continuum notion of causality in terms of
lightcone structure will emerge in a semiclassical approximation, can
instead replace it in deep quantum gravity regime; we have argued that
such a fundamental notion of causality can be implemented most
naturally in a sum-over-histories context; with the aim of
implementing this idea in the spin foam context, we have
linked the notion of causality with the
orientation of the 2-complex on which spin foam models are based, and
have identified the relevant data characterizing this orientation, and
found out
that all current spin foam models are
orientation-independent, i.e. defining amplitudes that trivial
functions of the orientation data; using the technology of evolution
kernels for quantum fields/particles on Lie groups and homogeneous
spaces, we have constructed a generalised version of spin foam models,
introducing an extra variable with the interpretation of proper time;
we have shown that different ranges of integration for this proper
time variable lead to different classes of spin foam models: one
corresponds to the usual ones, to be interpreted as the quantum
gravity analogue of the Hadamard function of QFT or equivalenty in a
canonical interpretation as a covariant  definition of the inner
product between quantum gravity states; the other is a new class of
models (one example of which having been constructed earlier in
\cite{lo}, and corresponds to the quantum gravity analogue of the
Feynman
propagator in QFT, i.e. a causal transition amplitude, a non-trivial
function of the orientation data, that implies a notion of \lq\lq
timeless ordering\rq\rq, based on purely group-theoretic methods, as
needed in Quantum Gravity; we have shown how the causal model is
manifestly related to simplicial gravity in the 4-dimensional
Lorentzian case. All these results hold true in full generality, for
the type of spin foam models considered, i.e. regardless of the
spacetime dimension and signature.  

In our opinion these results open quite a few lines of possible
further research. The causal/oriented models seem to solve the issue
of multiplicity of sectors present in all BF-type formulations of
quantum gravity \cite{review}, with different sectors isomorphic to
one another and
related by a change of orientation, all summed over in the path
integral quantization and thus interphering, leading to a discrepancy
of these models with the straightforward quantization of GR; in the
causal models it seems instead that a restriction to the GR sector is
achieved, as testified also by the final expression for the vertex
amplitudes in the 4-dimensional case shown above. This is likely to
have important consequencies for the
reconstruction of geometric quantities from the algebraic data, in
particular for the computation of expectation values of 3- and
4-volume operators, that are expected to give different results in
these new models as compared to those obtained in the literature \cite{perez}.        

The group field theory formulation of generalised and oriented models
should then be studied and work on this is indeed in progress, as a
natural step that would substantiate the particle picture extensively
used in their construction, and also to furnish a more complete
definition of the models, eliminating any dependence on a fixed
spacetime triangulation or 2-complex. This may turn out to be useful
also for defining a notion of \lq\lq positive and negative
energy\rq\rq sectors, also at the level of spin network states, based
on orientation/causal properties, in a timeless framework. Such a
result would also enforce the opinion, that we share, that the group
field theory is not an \lq\lq auxiliary\rq\rq formulation of spin foam
models and of quantum gravity, but the most fundamental formulation of
the theory, and as such should be the framework in which most problems
that this approach still faces are best tackled, including the issue
of semiclassical/continuum approximation. 

The proper time formalism we have developed could find interesting
applications in itself, with a first step being the study of the
relationship between our group theoretic proper time parameter and the
lapse function of canonical quantum gravity or the conformal factor
of the covariant formulation, and therefore with the
physical proper time for quantum gravity (both the lapse and the
conformal factor or spacetime volume element have used in the
literature as a proper time variable). Notice that, if such a
relationship is found, not only we would have strenghtened the
argument for the validity of our constuction and paved the way for
further developments, but we would have also obtained, as a side
result (!), the first definition to be found in the literature, to the
best of our knowledge, of a proper time expression for the action of
simplicial gravity, that reduces to the Regge action when the
additional variable is integrated out.

Another application of the proper time formalism could be the issue of
Wick rotation in spin foam models, and more gnerally in quantum
gravity, since it seems to be the right parameter in which to
analytically continue the amplitudes to define an \lq\lq
euclideanized\rq\rq model. 

Also, the generalised (proper time dependent) formulation of spin foam
models could represent a
new starting point for bridging the gap between spin foam models and
other approaches to quantum gravity in which proper time
plays a significant role, as for example the Lorentzian dynamical
triangulations, that have achieved recently important results
concerning the issue of continuum limit. 
The causal models seem indeed to be the at the point of
convergence of simplicial quantum gravity, dynamical triangulations
and causal sets, in addition to canonical loop quantum gravity, and
therefore represent the easiest context in which to analyse the
relationships between all these approaches. 

Finally, if the
orientation-independent models can be understood as defining the
matrix elements of the projector operator onto physical quantum
gravity states, then an intriguing possibility is to interpret the new
models as defining the matrix elements of an \lq\lq evolution
operator\rq\rq, whose property could be studied to understand for
example whether
a notion of unitary evolution is feasible in Quantum Gravity and the
\lq\lq scattering\rq\rq between quantum gravity perturbations, in
absence of an external time coordinate, but with a clearly identified
notion of causality.

\section{Acknowledgments}
I would like to thank gratefully Etera Livine and Florian Girelli for discussions and encouragement, and the organizers and staff of the DICE 2004 Workshop for a stimulating and truly enjoyable meeting.

\bibliography{apssamp}

\end{document}